\documentclass[twocolumn]{aastex62}

\def\lapp{\ifmmode\stackrel{<}{_{\sim}}\else$\stackrel{<}{_{\sim}}$\fi}
\def\gapp{\ifmmode\stackrel{>}{_{\sim}}\else$\stackrel{>}{_{\sim}}$\fi}
\usepackage{multirow}
\usepackage{color}
\usepackage{amsmath}
\usepackage{soul}
\usepackage{hyperref}
\usepackage{lineno}

\shorttitle{}
\shortauthors{}
\begin{document}

\title{X-ray timing studies of the low-field magnetar CXOU~J164710.2$-$455216}

\correspondingauthor{Hongjun An}
\email{hjan@cbnu.ac.kr}

\author{Hongjun An}
\affiliation{Department of Astronomy and Space Science,\\
Chungbuk National University, Cheongju, 28644, Republic of Korea}
\author{Robert Archibald}
\affiliation{Department of Astronomy and Astrophysics, \\
University of Toronto, 50 St. George Street, Toronto, ON M5S 3H4, Canada}

\begin{abstract}
        We report results of X-ray timing analyses for the low-field
magnetar CXOU~J164710.2$-$455216 which exhibited multiple outbursts.
We use data taken with {\it NICER}, {\it NuSTAR},
{\it Chandra}, and {\it Neil-Gehrels-Swift} telescopes between 2017 and 2018 when the source was in
an active state. We perform semi-phase-coherent timing analyses to measure the spin parameters
and a spin-inferred magnetic-field strength ($B_s$) of the magnetar.
Using a semi-phase-coherent method, we infer the magnetic field strengths
to be $3-4\times 10^{13}$\,G at the observation period ($\sim$MJD~58000), and
by comparing with previous frequency measurements (MJD~54000)
a long-term average value of $B_s$ is estimated to be $\approx4\times 10^{13}\rm \ G$.
So this analysis may add CXOU~J164710.2$-$455216 to the ranks of low-field magnetars.
The inferred characteristic age ($\tau_c$) is 1--2\,Myr which is smaller than the age
of Westerlund~1, so the magnetar's association with the star cluster is still secure.
For the low dipole field and the large age, recent multiple outbursts observed from the source are
hard to explain unless it has strong magnetic multipole components.
We also find timing anomalies around outburst epochs, which suggests that there may
be spin-down torque applied to the magnetar near the epochs as was proposed in magnetar models.
\end{abstract}


\section{Introduction}
\label{sec:intro}
	Magnetars are neutron stars with strong magnetic field typically greater
than $10^{14}\rm \ G$ \citep[][]{td95, td96}. They emit almost exclusively in the
X-ray band and so are observed in that band as pulsating sources. Their rotation
periods are in the relatively narrow range of 2--12\,sec, and the spin-inferred
dipole magnetic-field strengths at the surface $B_s \equiv 3.2\times 10^{19}\sqrt{P\dot P}$ are greater than
$10^{14}\rm \ G$ for most magnetars \citep[see][for more details]{ok14}.\footnote{See
the online magnetar catalog for general properties:
http://www.physics.mcgill.ca/$\sim$pulsar/magnetar/main.html}
Note that the dipole formula is only representative because $\dot P$ 
for magnetars is highly variable perhaps due to dynamical behavior of field lines.

	X-ray luminosity of many magnetars is greater than their spin-down power,
and they sometimes exhibit an outburst, a sudden increase in flux by orders of
magnitudes \citep[see][for reference]{mpm15,kb17}.
In addition, X-ray spectra of some magnetars show a turn-up
at $\sim$10\,keV \citep[e.g.,][]{khdc06}. As these are not often observed
in conventional rotation-powered pulsars (RPPs) with
typical $10^{12}-10^{13}\rm \ G$ fields, it was suggested that the strong magnetic field
of magnetars should play important roles inside \citep[e.g.,][]{pp11,pr12}
or outside the star \citep[e.g.,][]{tlk02,bel13} to give magnetars the observational
properties.

	However, discovery of low-field
magnetars \citep[e.g., SGR~0418$+$5729 and Swift~J1822.3$-$1606;][]{reti+10,skc14}
and RPPs with outbursts \citep[e.g., PSR~J1846$-$0258 and PSR~J1119$-$6127;][]{gggk+08,akts18}
suggests that there should be other important factors (besides dipole magnetic fields)
needed to explain the observational properties.
Promising candidates are magnetic-multipole components \citep[e.g.,][]{pp11, aec11,tzpe+11,cvpp19};
hints of these were found in some magnetars \citep[SGR~0418+5729, 1E~1048.1$-$5937;][]{temt+13,akbk+14}
where local magnetic loops were suggested based on possible cyclotron-line features in their spectra.
Nevertheless, the dipole fields are still important, and it was suggested that outburst rates
of magnetars depend sensitively on the dipole magnetic-field strength (for a given age and multipole strength)
in magneto-thermal evolution models \citep[e.g.,][]{vrpp+13}.
Therefore, accurately determining dipole magnetic-field strengths of magnetars with outbursts
will be particularly useful to provide further inputs to the models, which can be
done with precision timing analyses.
Furthermore, measurements of timing properties of a magnetar in outburst
can be used to test magnetar models with twisted fields \citep[e.g.,][]{tlk02, b09} which predict
that timing properties of a magnetar may change before and after an outburst 
due to enhanced spin-down torque.

	CXOU~J164710.2$-$455216 (J1647 hereafter) is an old ($\tau_c$$\ge$400\,kyr)
10.6-s magnetar possibly associated with the massive star cluster Westerlund~1 \citep[][]{mccd+06}.
It exhibited multiple outbursts \citep[e.g.,][]{icdm+07,akac13,brtp+19}, which makes this source
particularly useful for studying outburst relaxation \citep[e.g.,][]{ack18}
and magneto-thermal evolution \citep[e.g.,][]{pp11,vrpp+13}. However, the value of
the spin-inferred dipolar magnetic-field strength of J1647 is controversial.
A large value $\sim$$10^{14}$\,G was suggested based on data taken after 2006
and/or 2011 outbursts \citep[][]{icdm+07,wkga11,riep+14}, but
\citet[][]{akac13} inferred a time-average value of $B_s \le 7\times 10^{13}$\,G with
90\% confidence using the same data. While the inferred range of $B_s$
is not very large, it is important to measure $B_s$ accurately for this magnetar
as it may be a low-field magnetar with multiple outbursts, contrary to what
we expect in the standard evolutionary model \citep[e.g.,][]{pp11}.

	In this letter, we report our results of timing analyses for the possible
low-field magnetar J1647 performed using X-ray data taken between 2017 and 2018
when the source was active. Spectral properties of the source
during the time were previously reported in details \citep[][]{brtp+19},
so here we focus on timing analyses. We present our analysis results
and compare them with previous ones in Section~\ref{sec:sec2},
and then discuss and conclude in Section~\ref{sec:sec3}.

\section{Observational Data and Analysis}
\label{sec:sec2}

\newcommand{\marka}{\tablenotemark{a}}
\newcommand{\markb}{\tablenotemark{b}}
\begin{table}[t]
\vspace{-0.0in}
\begin{center}
\caption{Observational data used in this work}
\label{ta:ta1}
\vspace{-0.05in}
\scriptsize{
\begin{tabular}{cccc} \hline\hline
Observatory        & Obs. ids  & $N_{\rm obs.}$ & $N_{\rm pass}$\marka \\  \hline
{\it NICER}        & 0020350101--1020350192     &  82  & 39 \\
{\it NuSTAR}       & 80201050002, 4, 6, 8       &  4   &  4 \\
{\it Chandra}      & 19135--8, 20976      &  5   &  5 \\
{\it Swift}\markb  & 00030806064, 7, 8          &  3   &  2 \\ \hline
\end{tabular}}
\end{center}
\vspace{-0.5 mm}
\footnotesize{
$^{\rm a}${Number of observations which pass our criteria: $H>25$ and net exposure $>$200\,s after removing flares (see text).}\\
$^{\rm b}${Window-timing (WT) mode only.}
}\\
\end{table}

\subsection{Data reduction}
\label{sec:sec2_1}
	We use X-ray data taken with  {\it NICER} \citep[][]{gao12}, {\it NuSTAR}, {\it Chandra},
and {\it Swift} observatories between 2017 and 2018 (Table~\ref{ta:ta1}).
The {\it NICER}, {\it NuSTAR} and {\it Swift} data are processed
with the pipeline software for each observatory integrated in HEASOFT 6.25,
and the {\it Chandra} data are processed with {\tt chandra\_repro}
of CIAO 4.10 along with the most recent calibration database.
We use standard filters except for the {\it NuSTAR} data process for which
we use strict filters to remove enhanced
background near the South Atlantic Anomaly (SAA) passage
({\tt saamode=optimized} and {\tt tentacle=yes}); this turns out not to have
significant impact on the results below.

\subsection{X-ray timing analysis}
\label{sec:sec2_2}
	For timing analyses, we barycenter-correct the arrival
times using the {\it Chandra}-measured source position of
R.A.=$251.79242^\circ$ and decl.=$-45.871306^\circ$.
We extract events from the imaging data using circular
apertures of $R=30''$ and $2''$ circles for {\it NuSTAR} and {\it Chandra}, respectively.
It is very difficult to analyze {\it Swift}/PC data with a semi-phase-coherent method
as noted by \citet{brtp+19} because of the low timing resolution (2.5\,s), small
photon collecting area and short exposures. So we analyze the ``window-timing'' (WT)
mode data only. We extract source events in the {\it Swift}/WT data using $20''\times100''$ boxes.
We perform an initial timing analysis using the $H$ test \citep[][]{drs89} to
set an optimal energy range for each instrument and find that 1.2--5\,keV, 3--10\,keV,
0.5--5\,keV, and 1--10\,keV bands are optimal for {\it NICER}, {\it NuSTAR}, {\it Chandra},
and {\it Swift} data, respectively. We use these energy ranges in analyses below but the results
do not alter significantly if we change the energy bands slightly.

	For timing analyses to estimate $B$, high-quality pulse profiles need to be constructed.
However, during these observations the source is in an active state \citep[][]{brtp+19},
and there may be some low-level flares which can distort the pulse profiles.
In addition some background flares may also be problematic if there is any.
In particular, the {\it NICER} data are heavily contaminated by flares
but it is not clear whether or not these are from J1647;
we do not find contemporaneous flares in overlapping data taken by the other instruments.
Although the low-level flares from the magnetar can give us important information
on the outburst relaxation of the source, here we focus on the persistent behavior.
So we remove the flares whether or not they are from the source in order to measure the
``persistent'' pulse profiles accurately.
We construct light curves with 1-s and 10-s time scales, search them for time bins
which have larger counts ($>3\sigma$) than the time-average value,
and remove events in the time bins. We repeat this
process until there is no more high-count bin in the light curves.
There are very short flares ($\Delta t<0.1$\,s) in some {\it NICER} observations,
and we also remove these. We then visually inspect the cleaned light curves
and verify that there is no residual flare.
This process removes a significant amount of the {\it NICER} exposures, and so
unequal exposure among phase intervals and statistical fluctuation of background
are concerns. So we further require that the $H$ value for pulsation
should be greater than 25 and the net exposure should be larger than 200\,s ($\sim$20 rotations)
in each observation. This removes 43 {\it NICER} and 1 {\it Swift} observations.

\begin{figure}
\centering
\includegraphics[width=3.3 in]{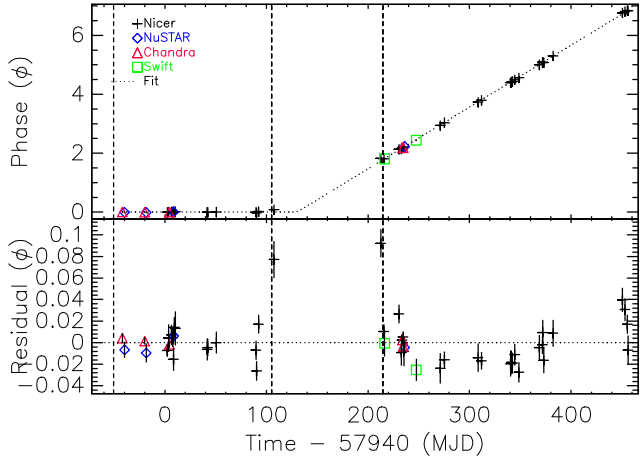}
\figcaption{Timing behavior of J1647 during the observations. Epochs of
outbursts are shown as vertical lines. Note that
there is rotation ambiguity between data sets before and after MJD~58100, so
the plot needs to be read with care.
A frequency change and a quadratic trend in the later data are clearly visible.
\label{fig:fig1}
}
\vspace{0mm}
\end{figure}

\begin{figure*}[ht]
\centering
\begin{tabular}{cc}
\includegraphics[width=3.3 in]{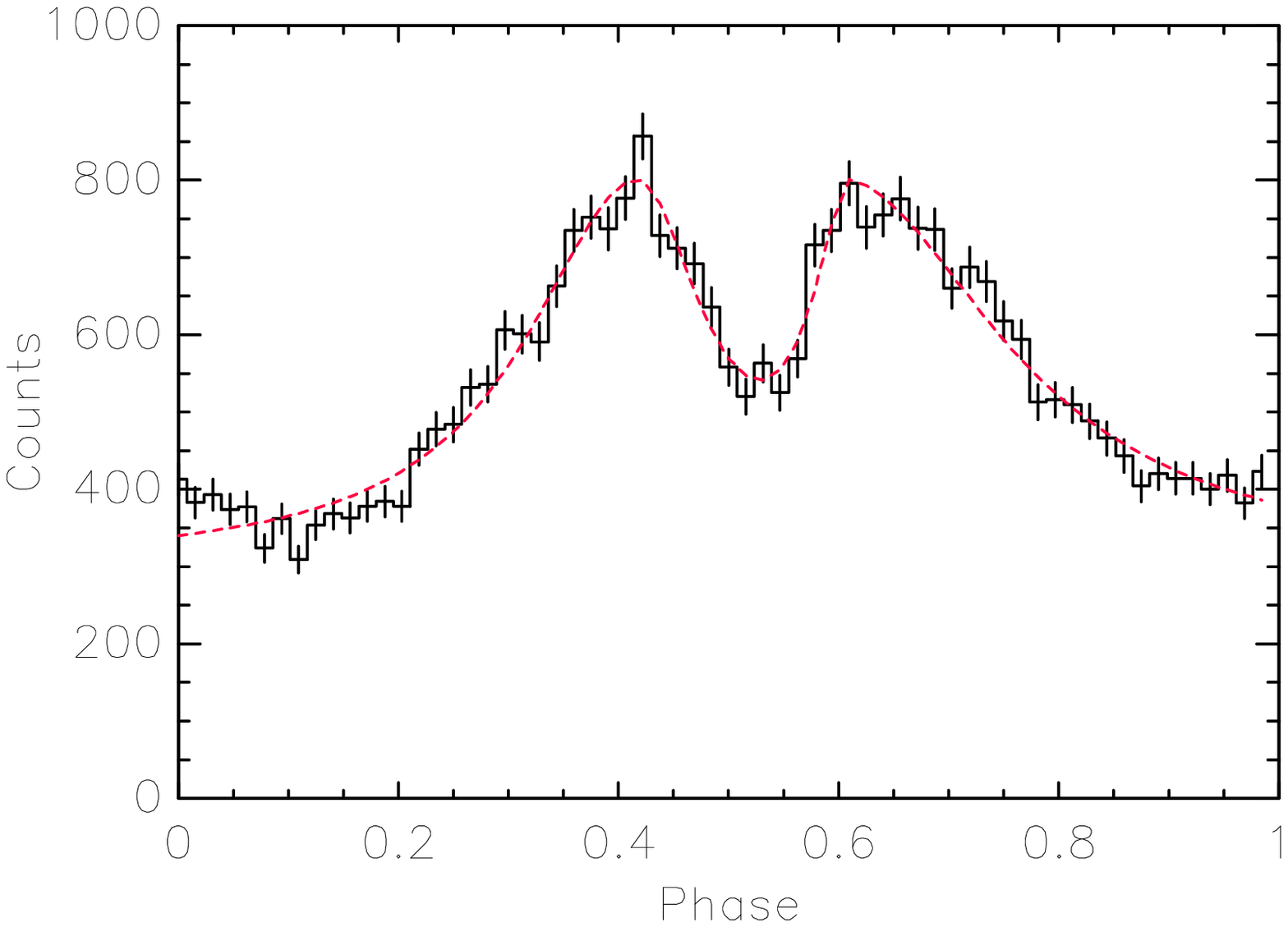} &
\includegraphics[width=3.3 in]{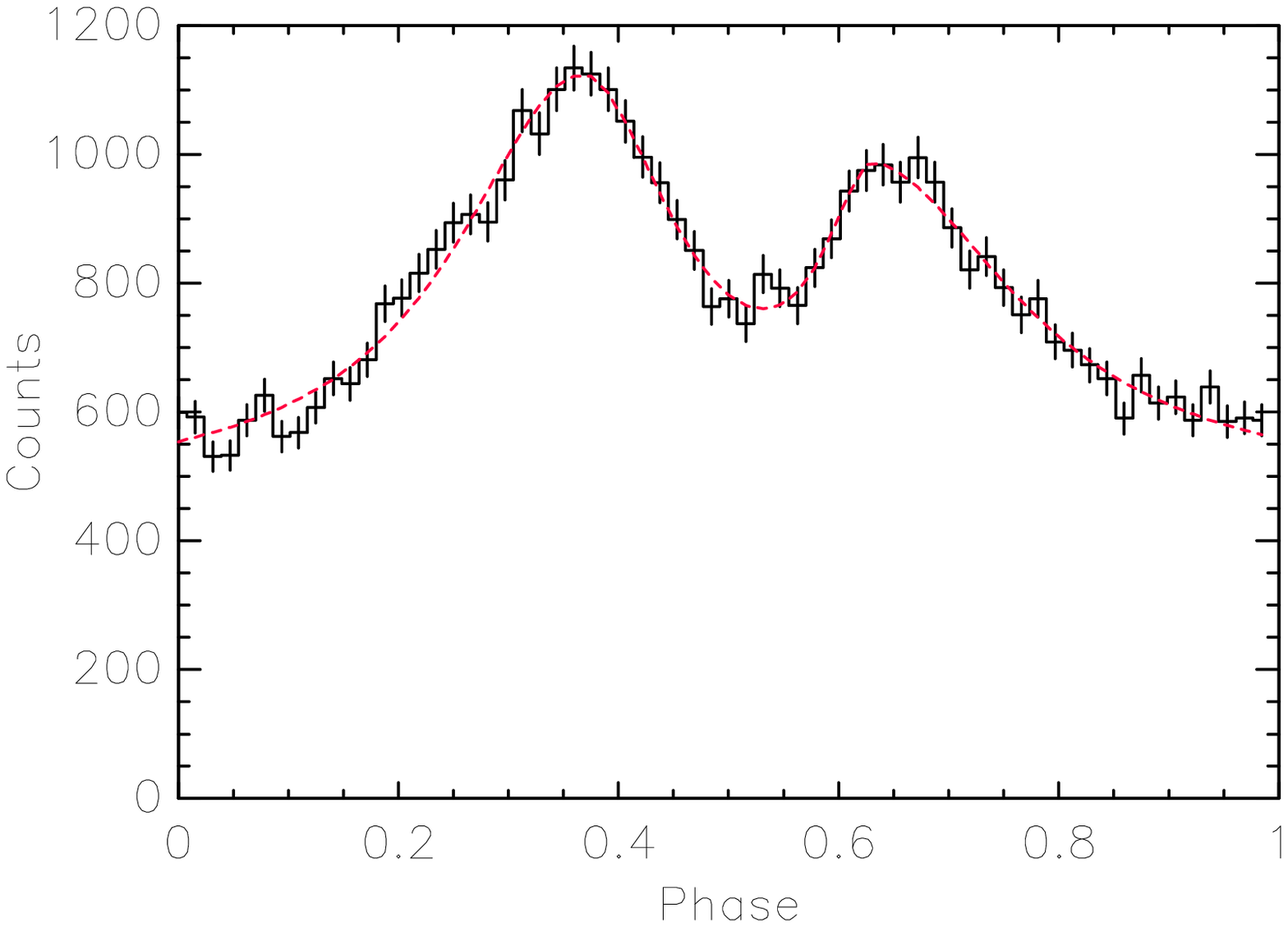} \\
\includegraphics[width=3.3 in]{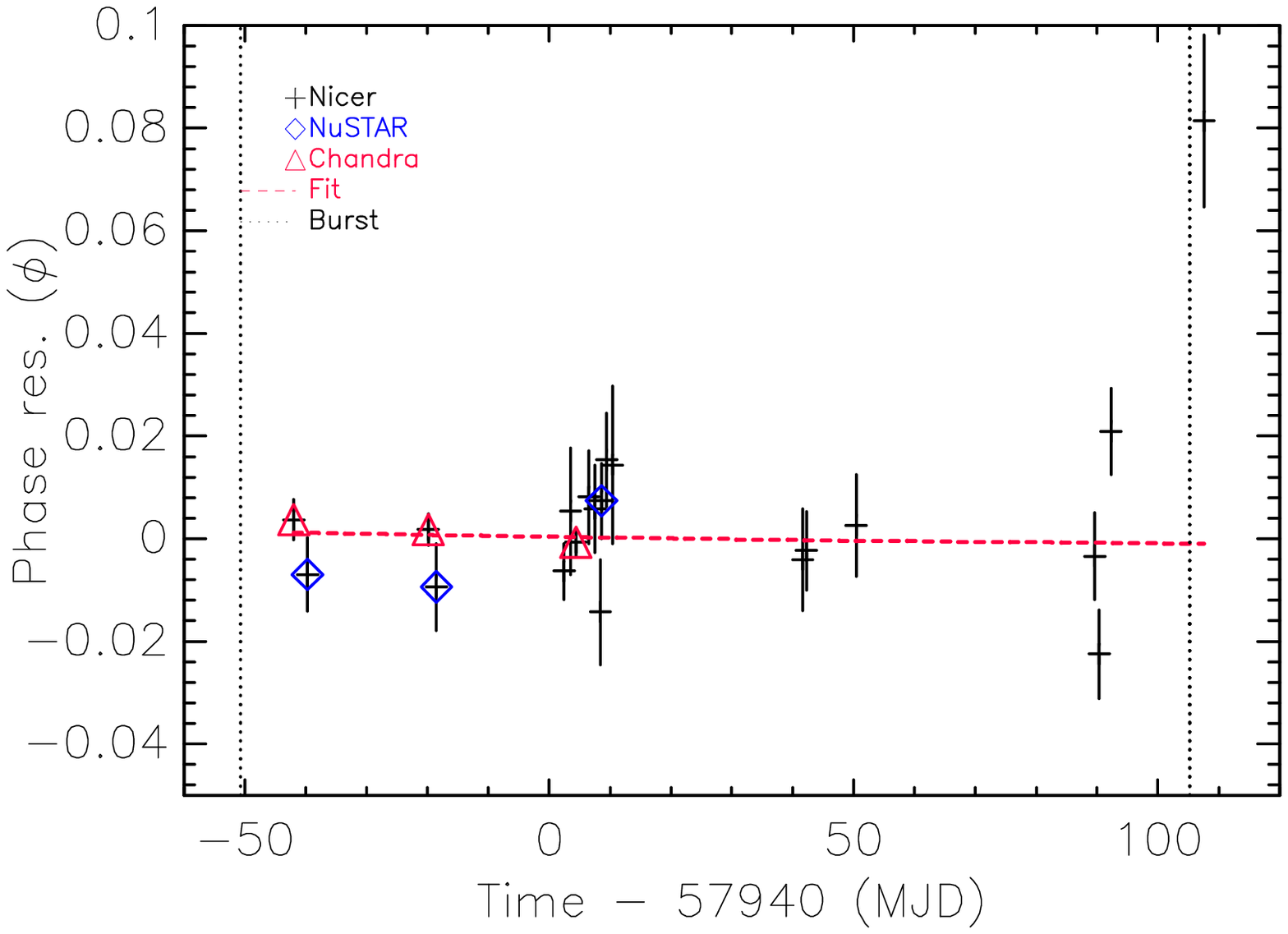} &
\includegraphics[width=3.3 in]{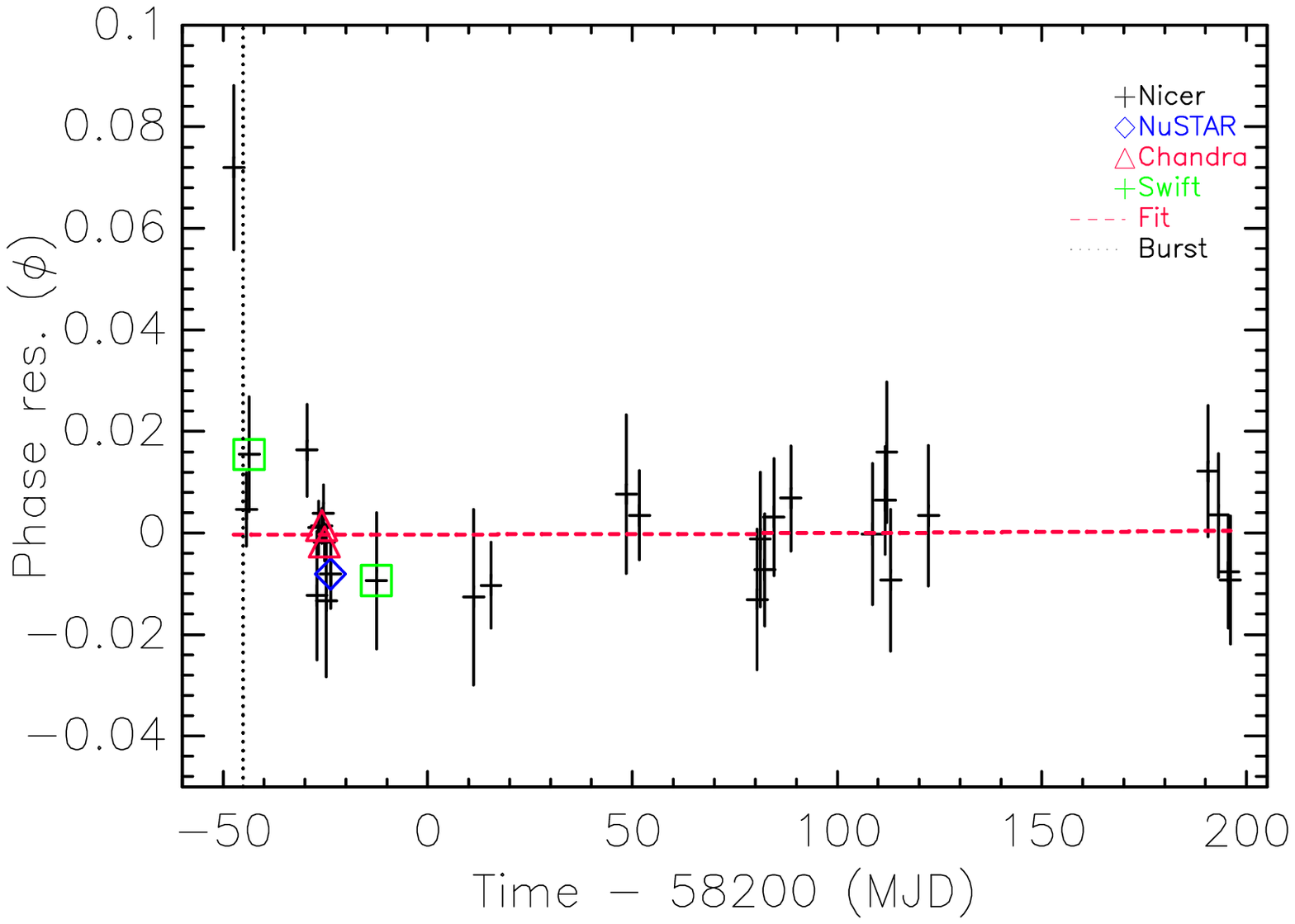} \\
\end{tabular}
\figcaption{Pulse-profile templates and timing residuals for segs.~1 (2017) and 2 (2018).
Figures in the left column are a pulse-profile template (top) and timing residuals (bottom) made with
the {\it NICER}, {\it Chandra}, and {\it NuSTAR} data for seg.~1,
and figures in the right column are a pulse-profile template and timing residuals for seg.~2.
Best-fit functions are also shown in red dashed lines, and epochs of outbursts are shown in vertical lines.
Note that the last point in the left panel and first three points in the right panel are excluded
in the fit.
\label{fig:fig2}
}
\vspace{0mm}
\end{figure*}

	Next, we measure the spin periods in the earlier {\it Chandra} and {\it NuSTAR}
data by searching for pulsations near the reported period of $P=$10.6106\,s using $H$ tests.
The pulsations are detected with high significance in these data, and we use these periods
(e.g., $P=10.6106(1)$ in {\it Chandra} Obs. ID 19135) as our starting point. We perform
a semi-phase-coherent timing analysis by folding the data on the period.
After iterating the analysis a few times by adjusting the frequency ($f$) and
its first time derivative ($\dot f$), the arrival phases align well until MJD~58047 when the phase
jumps (Fig.~\ref{fig:fig1}). The phase jump may imply an anti-glitch,
and so we check to see if $f$ at this epoch is significantly lower than the model
prediction (see below) using a $H$ test but the data are insufficient to discern
($\sim$1$\sigma$). In addition, the significance for
pulsation is not very high ($H$$\sim$25) and the pulse profile appears to be
different from the others.

	Nevertheless, we keep connecting the phases to the end of the data set,
and the results are shown in Figure~\ref{fig:fig1}. The phases after MJD~58152
do not align with the previous ones and there may be arbitrary phase wraps,
so the figure should be read with care; we make it to find frequencies to be
used as starting points in analyses below. Although there may be phase wraps between
the earlier and the later data, Figure~\ref{fig:fig1} shows a clear quadratic
trend in the later data (bottom). Since a single timing solution cannot be used throughout,
we divide the data into two segments: before (seg.~1) and after (seg.~2) the gap.

	We separately phase-connect the data in each segment and
construct pulse-profile templates. For the templates, we use observations in which
pulsations are detected with very high significance $H>70$.
The templates are fit with two asymmetric Lorentzian
functions, and we reanalyze the data with the template functions (Fig.~\ref{fig:fig2}).
The templates of segs.~1 and 2 differ slightly;
the second peak is relatively larger and
the separation between the peaks is smaller for seg.~1 than those in seg.~2;
the separations of the peaks are $\Delta\phi=0.195\pm0.008$ and $0.268\pm0.009$
for segs.~1 and 2, respectively.

	We fold the data in each observation to produce a pulse profile,
group the pulse profile to have at least 20 events per phase bin,
and fit the profile with the template function
by allowing the amplitudes and peak locations of the two Lorentzians
and the background level to vary.
Note that fitting both the locations and amplitudes of the peaks
is very important because the pulse profiles may vary with
time due to residual low-level flares (especially in the {\it NICER} data)
and/or an intrinsic variabilities.
For measuring arrival phases, we take the center position of the two peaks.
We measure the arrival phases for all the observations in each segment, fit the
phases with a quadratic function, and adjust $f$ and $\dot f$
to make the fit residuals flat; we find that higher time derivatives
are unnecessary in either data segments.
We update the profile template and shape parameters of the template functions,
and then repeat the above process until no more change
of $f$ and $\dot f$ is necessary. The final phase residuals are shown in Figure~\ref{fig:fig2}.
Note that for the {\it Chandra} analysis, we added a systematic
uncertainty corresponding to the timing resolution of 0.44\,s for the subarray
observations.

\newcommand{\markaa}{\tablenotemark{a}}
\begin{table*}[t]
\vspace{-0.0in}
\begin{center}
\caption{Comparison with previous frequency measurements}
\label{ta:ta2}
\vspace{-0.05in}
\scriptsize{
\begin{tabular}{clll|llllc} \hline\hline
Epoch     & $f$               & $\dot f$                & $B_s$           & $\dot f_{\rm 2017}$\markaa & $B_{s\rm, 2017}$\markaa   & $\dot f_{\rm 2018}$\markaa & $B_{s\rm, 2018}$\markaa  &  Ref. \\ 
          & $\rm s^{-1}$      & $10^{-15}\rm \ s^{-2}$  & $10^{13}$\,G    & $10^{-15}\rm \ s^{-2}$ & $10^{13}$\,G       & $10^{-15}\rm \ s^{-2}$ & $10^{13}$\,G           &            \\ \hline
53999.0   & 0.0942448896(18)  & $-8.2(6)$   & 10.0(2)   & $-1.009(5)$     & 3.51(1)                & $-1.651(5)$        & 4.494(7)               &     1      \\
53999.1   & 0.09424498(15)    & $>$$-3.6$   & $<$$7$    & $-1.3(4)$       & 4.0(7)                 & $-1.9(4)$          & 4.8(5)                 &     2      \\ 
53999.1   & 0.0942448814(4)   & $-8.633(9)$ & 10.277(5) & $-0.987(2)$     & 3.474(4)               & $-1.631(2)$        & 4.467(3)               &     3      \\
54008.0   & 0.0942448774(11)  & $-7.4(2)$   & 9.5(1)    & $-0.975(4)$     & 3.454(7)               & $-1.620(4)$        & 4.452(5)               &     4      \\
54008.0   & 0.0942448813(14)  & $-11.4(9)$  & 11.8(5)   & $-0.987(4)$     & 3.474(8)               & $-1.631(3)$        & 4.467(6)               &     4      \\
57940.0   & 0.0942445461(6)   & $-0.6(2)$   & 2.9(4)    & $\cdots$        & $\cdots$               & $\cdots$           & $\cdots$               & Seg.~1 of this work  \\
58160.0   & 0.0942442962(7)   & $-1.37(8)$  & 4.1(1)    & $\cdots$        & $\cdots$               & $\cdots$           & $\cdots$               & Seg.~2 of this work  \\ \hline
\end{tabular}}
\end{center}
\vspace{-0.5 mm}
\footnotesize{Refs. [1] \citet{icdm+07}, [2] \citet{akac13}, [3] \citet{riep+14}, [4] \citet{wkga11}.\\}
$^{\rm a}${Time-average $\dot f$ and magnetic field strengths estimated by comparing with our segs.~1 (2017) and 2 (2018) results.}\\
\end{table*}

	For the data in seg.~1 we measure the spin frequency and
its first derivative to be $f=0.0942445461(6)\rm \ s^{-1}$ and
$\dot f=-6.4\pm 1.9\times 10^{-16}\rm \ s^{-2}$ ($\chi^2/dof=26/17$),
implying $B_s=2.9\pm 0.4\times 10^{13}$\,G and $\tau_c=2.3$\,Myr;
the latter is smaller than the age of the Westerlund~1 cluster \citep[3.5--5\,Myr;][]{cncg05}.
The last data point in this segment is not used for deriving the timing solution
even though the point is shown in Figure~\ref{fig:fig2}; including
this point increases $B_s$ to $3.5\times 10^{13}$\,G but makes the fit significantly
worse ($\chi^2/dof=47/18$).

	After an outburst in 2018 \citep[MJD~58155;][]{brtp+19},
timing properties of the source changed significantly;
the frequency is smaller and $|\dot f|$ is
significantly larger than those measured in seg.~1.
In these data (seg.~2), the spin frequency and its first derivative are measured to be
$f=0.0942442962(7)\rm \ s^{-1}$ and
$\dot f=-1.37\pm 0.08\times 10^{-15}\rm \ s^{-2}$ ($\chi^2/dof=15/25$), 
implying $B_s=4.1\pm0.1\times 10^{13}$\,G and $\tau_c=1$\,Myr.
The measured spin parameters are presented in Table~\ref{ta:ta2}.
Note that we ignore the data points in MJD~58152--58157 around the outburst
epoch in this analysis because they might have been affected by the outburst; when we include
these points in the fit, $B_s$ does not change significantly but the fit becomes worse ($\chi^2/dof=34/28$).

\subsection{Comparison with previous results}
\label{sec:sec2_3}

	Although we are able to estimate $B_s$ in the time period of 2017--2018,
this estimation may be biased by the outburst activities and/or undetected
long-term timing noise which might have changed the frequency
derivative \citep[e.g.,][]{scsr+17}. Then our estimation above may not be accurate,
perhaps higher than the actual value if the measurements are affected by the radiative activities.
However long-term $\dot f$ estimated with frequency differencing
might have been less affected by timing anomalies
because of the long baseline ($\sim$11 years), and so
estimating a long-term averaged $B_s$ using frequency differencing
may provide an independent check.
We do this by comparing our results with previous measurements made at the
reference epoch MJD~54000 \citep[][]{icdm+07,wkga11,akac13,riep+14}.

	In Table~\ref{ta:ta2}, we summarize previous spin-parameter measurements. While the
previous measurements are all done near the same epoch, the results differ significantly
because of different assumptions made in those works. However, the difference is small
($\Delta f\approx10^{-7}\rm \ s^{-1}$), so errors in estimating
$B_s \equiv 3.2\times 10^{19}\sqrt{P\dot P}$ would not be large.
Note that $\dot f$ appears to have changed dramatically since the previous measurements
which are probably affected by a putative glitch and its recovery
in 2006 as noted by \citet{wkga11} and \citet{akac13}.
This again justifies the time-differencing measurement.
We compare the previous $f$ values with our results and measure time-averaged $\dot f$
to estimate time-averaged $B_s$ for J1647. These values are shown in Table~\ref{ta:ta2}
and typically $B_s\approx 4\times 10^{13}$\,G.

\section{Discussion and Conclusions}
\label{sec:sec3}

	We analyzed X-ray data taken with {\it NICER}, {\it NuSTAR}, {\it Chandra} and
{\it Neil-Gehrels-Swift} observatories in 2017--2018 to measure timing properties of the magnetar J1647.
We found that the magnetar's spin properties changed significantly after the
outburst in 2018. We therefore split the data into two segments, and found that
the magnetic field strengths were low $B_s$$\sim$$3\times 10^{13}$ and $\sim$$4\times 10^{13}$\,G
in 2017 and 2018, respectively.
While these values may be biased because of `undetected' timing anomalies associated with
the magnetar's activities or timing noise, long-term time-averaged $B_s$
which provides an independent estimation is also low $B_s\approx 4\times 10^{13}$\,G.

	There were three major bursts in the data we analyzed \citep[MJDs~57889, 58045, and 58155;][]{brtp+19}.
For the later two, the source was observed with {\it NICER} within $\pm$2\,days
and so measuring timing properties near the outbursts were possible
although we ignored them when estimating $B_s$ above.
For the one measured at MJD~58047 (2 days after the activity),
the sudden shift in phase, if real, may be due to an anti-glitch \citep[e.g.,][]{akng+14}.
However a frequency measurement does not confirm this due to large uncertainty.
Furthermore,
the pulse profile at this epoch is distorted with one of the peaks not being clearly visible,
which may be due to increased constant emission, contamination from low-level flares, or
intrinsic (temporary) change of the profile. So the phase shift in this case is rather uncertain.

	For the outburst at MJD~58155, a {\it NICER} observation was taken $\sim$2\,days
`before'. For this, the pulse profile appears
to be normal and the detection significance is very high ($H\approx 52$). So this
shift seems to be real.
{\it Swift} monitoring data \citep[][]{brtp+19} do not cover the same period,
so it is not clear whether or not this shift is associated with a spectral change.
Nevertheless, the arrival time of the shifted pulse ($\sim$2\,days `before' the outburst)
was later than expected from the rest of the data in the segment,
which may imply that the star was spinning slower at the time (implying a glitch) although
it is hard to tell conclusively without any measurement right before (i.e., a precise measurement of $f$).
In the next observation at MJD~58156 the phase shift was
recovered quickly ($\sim$1\,day `after' the outburst; Fig.~\ref{fig:fig2}) perhaps
by an enhanced spin-up (i.e., glitch). Theoretically this may be explained by enhanced spin-down torque
due to gradual buildup of magnetic twist before magnetar outbursts and a
temporary spin-up (i.e., glitch) after releasing magnetic energy of
the twisted fields by an outburst, as proposed
in twisted-field magnetar models \citep[e.g.,][]{tlk02, b09, cvpp19}.
The observations are sparse, so we were not able to measure the evolution of
the spin-down torque before and after the outburst. More detailed measurements of spin properties of magnetars
near outburst epochs can help to improve the models further.

	The pulse profiles in the earlier (seg.~1) and the later (seg.~2) data differ;
the second peak is relatively larger and the separation
between the peaks is smaller in the earlier ones (e.g., Fig.~\ref{fig:fig2}).
While the change is not as dramatic as those seen after the 2006 (single to three peaks)
or the 2011 (single to two peaks) outburst \citep[e.g.,][]{riep+14},
the change seems to be real as it can be seen by
comparing individual observations with large statistics (e.g., {\it Chandra} data).
The separation of the peaks in the first observation of seg.~2 (2\,days before the 2018 outburst)
is smaller $\Delta \phi=0.20\pm0.03$
than those in the same segment but similar to those in seg.~1.
This suggests that the change might have occurred near the outburst at MJD~58155.
Moreover, emission (dominated by the low-energy band) of J1647 during the observation period
is well described by a thermal model, and the size of the emitting region
increased after the 2018 outburst \citep[][]{brtp+19}.
Changes of magnetars' emission properties after an outburst are
expected in outburst relaxation models \citep[e.g.,][]{bel13,cvpp19} which predict
that emission at the stellar surface (the magnetic footprint on the star) could change
after an outburst by bombardment of return currents in differently configured magnetic fields.
However, the models are yet qualitative, and the statistics and cadence of the
observations are not sufficient to measure the evolution of the temporal and spectral
properties in details. Further theoretical works to make quantitative interpretation
of spectral/temporal evolution and more observations near/after magnetar outbursts can help
to understand mechanisms of outburst relaxation.

	We estimated the magnetic-field strength of J1647 to be
$\approx 4\times 10^{13}$\,G. Note that previous estimations of $|\dot f|$ and so $B_s$
\citep[$\sim$$10^{14}$\,G;][]{icdm+07,wkga11,riep+14} are large but they may be biased by a putative glitch
and its recovery at the time as noted by \citet{wkga11} and \citet{akac13}.
Although more data are needed to measure the true ``baseline'' $B_s$ (e.g., without timing-noise effects),
our measurements suggest that J1647 may add to the list of low-field magnetars.
The source has shown at least five outbursts since the first one detected in 2006.
The low dipole-field strength, large characteristic age
and frequent activities suggest that multipole components should
be strong in J1647 \citep[e.g.,][]{pp11,vrpp+13};
its complex pulse profiles (1--3 peaks) previously observed after outbursts
may be related to the multipole components. Or do magnetar outbursts cluster in time so that
a magnetar is more likely to outburst over a certain period of time for a given long-term
average outburst rate? Then, maybe we are observing clusters of outbursts of J1647 while its
long-term average rate is actually very low for the low dipole field and large age.
The low field strength and multiple outbursts make J1647 a particularly intriguing source
for study of magnetar evolution, and future observational and theoretical works may give us new insights
into magnetar physics.

\bigskip
\bigskip

\acknowledgments

We thank the anonymous referee for careful reading of the paper and
insightful comments.
This research was supported by Basic Science Research Program through
the National Research Foundation of Korea (NRF)
funded by the Ministry of Science, ICT \& Future Planning (NRF-2017R1C1B2004566).

\vspace{5mm}
\facilities{NICER, NuSTAR, CXO, Swift}
\software{HEAsoft (v6.25; HEASARC 2014), CIAO \citep[v4.10; ][]{fmab+06}}

\bibliographystyle{apj}
\bibliography{MAGNETAR,GBINARY,BLLacs,PSRBINARY,PWN,STATISTICS,FERMIBASE,COMPUTING,INSTRUMENT,ABSORB}

\end{document}